\documentclass[
prd
,showpacs,amssymb,superscriptaddress,aps
]{revtex4}
\usepackage{graphicx}
\usepackage{color}
\input{epsf}

\usepackage{amsmath,amssymb}
\usepackage{bm}
\usepackage{times}
\usepackage{ulem}

\usepackage{hyperref}
\hypersetup{
    colorlinks=true,
    linkcolor=blue,
    filecolor=magenta,      
    urlcolor=cyan,
    citecolor=blue
}
\newcommand{\dalm}{\kern1pt\vbox{\hrule height 0.9pt\hbox{\vrule width 0.9pt
\hskip 2.5pt\vbox{\vskip 5.5pt}\hskip 3pt\vrule width 0.3pt}\hrule height 0.3pt}
\kern1pt}

\begin{document}



\title{Universality in supernova gravitational waves with proto-neutron star properties}

\author{Hajime Sotani}
\email{sotani@yukawa.kyoto-u.ac.jp}
\affiliation{Astrophysical Big Bang Laboratory, RIKEN, Saitama 351-0198, Japan}
\affiliation{Interdisciplinary Theoretical \& Mathematical Science Program (iTHEMS), RIKEN, Saitama 351-0198, Japan}

\author{Bernhard M\"{u}ller}
\affiliation{Monash Centre for Astrophysics, School of Physics and Astronomy, 11 College Walk, Monash University, Clayton VIC 3800, Australia}

\author{Tomoya Takiwaki}
\affiliation{Division of Science, National Astronomical Observatory of Japan, 2-21-1 Osawa, Mitaka, Tokyo 181-8588, Japan}
\affiliation{Center for Computational Astrophysics, National Astronomical Observatory of Japan, 2-21-1 Osawa, Mitaka, Tokyo 181-8588, Japan}

\date{\today}

\begin{abstract}
Gravitational wave signals from core-collapse supernovae are one of the important observables for extracting the information of dense matter. To extract the properties of proto-neutron stars produced via core-collapse supernovae by asteroseismology, we perform a linear perturbation analysis using data obtained from two-dimensional numerical simulations. We employ 12 and 20 solar-mass progenitors and compare two different treatments of gravity. One is a general relativistic one with a conformal flatness condition and the other is an effective gravitational potential mimicking the Tolman-Oppenheimer-Volkoff solution. We discuss how the frequencies of the proto-neutron star oscillations corresponding to the gravitational wave signals in the simulations depend on the proto-neutron star properties. In our models, we find that the gravitational wave frequencies of the proto-neutron stars determined with the Cowling approximation can be expressed to very good approximation as a function of the proto-neutron star average density almost independently of the progenitor mass, treatment of gravity in the simulations, and the interpolations in the simulations. On the other hand, if one considers the gravitational wave frequencies as a function of the surface gravity of proto-neutron stars, such a relation appears sensitive to the treatment of gravity and other numerical details in the simulations. Thus, the average density of proto-neutron stars seems more suitable for universally expressing the supernova gravitational wave frequencies, instead of the surface gravity.
\end{abstract}

\pacs{04.40.Dg, 97.10.Sj, 04.30.-w}
%
\maketitle


\section{Introduction}
\label{sec:I}

Direct detection of gravitational waves from binary black holes and neutron stars have opened the door to a new era of observational astronomy. Gravitational waves have now become one of the important observables to see (or hear) from a compact object, complementing astronomical information together with electromagnetic waves and neutrinos. In the gravitational wave event due to a merger of the binary neutron stars, GW170817 \cite{GW6}, the electromagnetic waves have also been observed from the same object as counterparts \cite{EM}. The fourth observing (O4) run by the LIGO-Virgo-KAGRA collaboration has been operating since May 2023 and promises to deliver a multitude of new detections of merger events. Next to mergers of compact binary systems, core-collapse supernovae are among the most promising sources of gravitational waves. Since supernovae are less aspherical than merging binary systems, the energy of gravitational waves from the supernovae is weaker by comparison and limits the achievable detection distance. Because of this greater difficulty in the detection of supernova gravitational waves, we have to gear up for such an interesting event with thoroughgoing preparation to optimize detectability and the extraction of physical information from a prospective signal.

Studies about supernova gravitational wave signals have mainly been done via numerical simulations, e.g.,~\cite{Murphy09,MJM2013,Ott13,CDAF2013,Yakunin15,KKT2016,Andresen16,Richers2017,Takiwaki2017,OC2018,RMBVN19,VBR2019,andresen2019,PM2020,vartanyan20,Andersen21,andresen21,Vartanyan23,Jakobus23,Bruel23}. These studies showed the existence of time-dependent gravitational wave signals (ramp-up signals), whose frequencies increase from a few hundred Hertz up to the kHz range within $\sim 1$ second after core bounce. This signal was originally considered to correspond to the Brunt-V\"{a}is\"{a}l\"{a} frequency (or the surface gravity ($g$-) mode) at the proto-neutron star surface \cite{MJM2013,CDAF2013}. More precise analysis has identified the dominant frequency as that of fundamental ($f$-) mode oscillations (or $g$-mode sometimes although the mode classification may be different from the standard one) of proto-neutron stars \citep[e.g.,][]{MRBV2018,SKTK2019,ST2020b,TCPF2018,TCPOF2019a,TCPOF2019b,STT2021,Bizouard21,mori23,wolfe23,rodriguez23}.  In addition to the ramp-up signal, depending on the EOS, the excitation of gravitational waves, whose frequencies are around 100 Hz, is also reported, which is considered as a result of standing accretion-shock instability (SASI) outside the proto-neutron star \cite{KKT2016,Andresen16,andresen2019,andresen21,vartanyan20,Kawahara18,Takeda21,Lin23}. Furthermore, there may be secondary signals, whose frequencies are decreasing with time \cite{MRBV2018,Kawahara18,Jakobus23}, from the $g_1$ oscillation mode of the proto-neutron star, whose time-dependent behavior is strongly associated with the Brunt-V\"{a}is\"{a}l\"{a} frequency below the proto-neutron star convection zone \cite{ST2020b,Jakobus23}. Linear analysis has proved extremely important for identifying the origin of the gravitational wave signals appearing in the numerical simulations.

The linear analysis of oscillations in compact objects has been used for cold neutron stars since early times \cite{KS1999}. The spectrum of oscillation modes encodes abundant information about the neutron star structure and microphysics. Thus, one can extract the corresponding physics as an inverse problem, if one can identify the observed frequencies with specific modes \cite{AK1996,AK1998}. This technique is known as neutron star asteroseismology, which is analogous to the use of seismology on Earth and helioseismology on the Sun to probe inside of these objects using the oscillation patterns. Even for compact objects, this technique is powerful for extracting physical information. For example, the neutron star properties are constrained by identifying the quasi-periodic oscillations observed in the magnetar giant flares with the crustal torsional oscillations, e.g., \cite{GNHL2011,SNIO2012,SIO2016,SKS23}. Similarly, if one observes the gravitational waves from compact objects, one may extract the information of neutron star mass, radius, and equation of state (EOS) for neutron star matter, e.g., \cite{AK1996,AK1998,STM2001,SH2003,SYMT2011,PA2012,DGKK2013,Sotani2020}.

In principle, the technique of asteroseismology can also be adapted to proto-neutron stars formed in core-collapse supernovae. However, compared to the case of cold neutron stars, the asteroseismology of proto-neutron stars is more complicated. Importantly, it is more difficult to obtain the proto-neutron background structure from which oscillation modes can be computed by means of linear analysis.
Cold (zero temperature) neutron star models can be constructed simply with the relation between the density and pressure, i.e., the EOS with zero temperature, while one needs the information of the radial profiles of the electron fraction and entropy per baryon as well as the pressure and density to construct the proto-neutron star models. The requisite radial profiles need to be obtained from dynamical core-collapse supernova simulations. One can then determine the specific oscillation modes of proto-neutron stars using the simulation data and linear perturbation theory. The linear analysis can be done with any background models in principle, but up to now, several groups have simply done the asteroseismology on the spherically symmetric (1D) proto-neutron star models, obtained either as spherical averages from multi-dimensional simulations or from 1D simulations \citep[e.g.,][]{MRBV2018,SKTK2019,ST2020b,TCPF2018,TCPOF2019a,TCPOF2019b,STT2021,FMP2003,Burgio2011,FKAO2015,ST2016,Camelio17,SKTK2017,SS2019,WS2019,Mezzacappa20,ST2020a,ST2020c}.

Since proto-neutron stars are not surrounded by vacuum, but by a dense accretion flow, the choice of boundary conditions in the linear oscillation problem is non-trivial. Broadly speaking, two different approaches have been mainly adopted by different groups. The difference between them concerns the domain where one performs the linear analysis. The first approach follows the usual practice in asteroseismology by considering the dense ($\gtrsim 10^{11}\,\mathrm{g}\, \mathrm{cm}^{-3}$) hydrostatic central region as a proto-neutron star background model, and imposes the boundary condition that the Lagrangian perturbation of pressure should be zero at ``the proto-neutron star surface''. With this approach, one can classify the resultant oscillation frequencies into specific eigenmodes by definition because the problem to solve is mathematically the same as in standard asteroseismology, but one faces uncertainties about how to define the position of the proto-neutron star surface. Nevertheless, the frequencies of at least the $f$- and $g_1$-modes, which we focus on in this study, seem to be almost independent of the selection of the surface density \cite{MRBV2018,ST2020b}. In this study, we adopt this approach with the surface density being $10^{11} \,\mathrm{g}\,\mathrm{cm}^{-3}$. The second approach considers the whole region inside the shock radius. Because of the presence of matter outside the proto-neutron star, this approach is plausible, but the boundary condition imposed at the shock radius is that the radial displacement should be zero \cite{TCPF2018,TCPOF2019a,TCPOF2019b}, which is a completely different problem from the usual asteroseismology. So, one has to newly develop an appropriate classification of the resultant oscillation modes. Furthermore, the background structure cannot be treated as hydrostatic any longer if the post-shock region is included. Regardless of the technical approach, however, the ramp-up signals appearing in the numerical simulations can be confidently identified with the $f$-mode (or the ${}^2g_2$-mode in the classification with the second approach).

In the event of a gravitational wave observation from a supernova, one does not know many of the relevant physical parameters for constructing detailed supernova models to match gravitational waveforms, such as the progenitor mass and EOS for dense matter. Thus, realistically, detailed simulations cannot scan the relevant high-dimensional parameter space to provide enough template waveforms for parameter inference. For this reason, identifying ``universal'' relations between the gravitational wave frequencies and proto-neutron star properties that are almost independent of the supernova models is useful and critical for extracting physical information from prospective gravitational wave signals. Some universal relations have already been proposed in the literature for expressing the $f$-mode gravitational wave frequencies as a function of proto-neutron star average density, $M_{\rm PNS}/R_{\rm PNS}^3$, or surface gravity, $M_{\rm PNS}/R_{\rm PNS}^2$, and hence ultimately as a function proto-neutron star mass $M_{\rm PNS}$ and radius $R_{\rm PNS}$ \cite{TCPOF2019b,STT2021}. 
We note that universal relations for cold neutron stars have also been discussed sometimes in the literature, e.g., \cite{SK21,RS23,PCA23}.

Nevertheless, the proposed universal relations sometimes do not seem to capture numerical simulations well  \citep[e.g.,][]{wolfe23}. This may be due to differences in the treatment of gravity, or other details of the numerical schemes and physical assumptions in different simulations. In this study, we will therefore investigate which physical quantities can be used to construct universal relations for supernova gravitational wave frequencies by examining their dependence on different treatments of gravity and numerical methods.

This paper is organized as follows. In Sec.~\ref{sec:PNSmodel}, we describe the supernova simulations and resultant PNS models considered in this study. In Sec.~\ref{sec:GW}, we calculate the eigenfrequencies of gravitational waves from the PNS within the Cowling approximation, compare the gravitational wave signals appearing in the numerical simulations to frequencies of the proto-neutron star oscillations, and discuss the universal relations for the gravitational wave frequencies. We present our conclusions in Sec.~\ref{sec:Conclusion}. Unless otherwise stated, we adopt geometric units in the following, $c=G=1$, where $c$ denotes the speed of light, and the metric signature is $(-,+,+,+)$.

\section{Background models}
\label{sec:PNSmodel}

To discuss the connection between the time-frequency structure of gravitational wave signals, proto-neutron star oscillation modes, and proto-neutron star properties, multidimensional simulations of core-collapse supernovae are required to provide input for linear mode analysis. In this section, we describe the supernova simulations and the resultant proto-neutron star models considered in this study.

\subsection{Supernova models}
\label{sec:SN}

In this study, we perform two-dimensional (2D) simulations with the $12M_\odot$ and $20M_\odot$ progenitor models from Ref.~\cite{WH2007}. Hereafter we refer to these two models as S12 and S20, respectively. The simulations are performed with the SFHo EOS \cite{SFHo}. In order to investigate how the gravitational wave spectrograms and the neutron star mode frequencies depend on the treatment of gravity in the simulations, we perform simulations with two different treatments of gravity, i.e., with an effective relativistic potential (effective GR) and relativistic (GR) gravity in the conformally flat approximation.

The effective GR is done with the Newtonian hydrodynamic simulation, approximately taking into account the general relativistic effect (especially with case A in \cite{Marek06}). In this study, as in our previous studies \cite{ST2020b,STT2021} and the other studies, e.g.,  \cite{takiwaki2016,oconnor2018,kotake2018,nakamura2019,sasaki2019,zaizen2019}, the numerical simulations have been done with \textsc{3DnSNe} code \cite{takiwaki2016,matsumoto20,matsumoto22}. In particular, the simulations in this study have been performed with 5th-order interpolation \cite{matsumoto22} with Harten-Lax-van Leer discontinuities (HLLD) \cite{matsumoto20,miyoshi05}, which is a different interpolation from that adopted in our previous studies, e.g., \cite{ST2020b,STT2021}. We use monopole approximation ignoring multi-pole expansion of the gravitational potential \cite{wongwathanarat10}. The spatial range of the computational domain is within $r< 5,000$ km in radius, covered by 512 non-uniform zones. The polar grid in the spherical coordinate is in the range from 0 to $\pi$, covered by 128 zones.

Secondly, we conduct a general relativistic simulation using the extended conformal flatness condition  (xCFC) \cite{cordero2009} for the space-time metric. We use the \textsc{CoCoNuT-FMT} code, which combines the xCFC approximation of the metric with a finite-volume hydrodynamics solver using 6th-order piecewise parabolic reconstruction \cite{colella2008} and the HLLC Riemann solver \cite{mignone2006}, and treats neutrino with the fast-multigroup transport (FMT) method \citep{mueller2015}. The \textsc{CoCoNuT-FMT} simulation uses spherical polar coordinates with a non-equidistant radial grid of 550 zones extending from the origin to a radius of $10^{10}\,\mathrm{cm}$ and 128 zones in latitude. \textsc{CoCoNuT-FMT} can compute the multi-dimensional space-time metric using a multipole expansion for solving the non-linear Poisson equations in the elliptic xCFC system, or apply the monopole approximation. For comparison with the linear mode analysis in the Cowling approximation (see also Sec.~\ref{sec:GW}), dynamical perturbations in the quadrupole component of the metric functions should be neglected in the simulations, i.e., the monopole approximation is appropriate.

Furthermore, the gravitational wave emission and mode frequencies in the simulations may also depend on other details of the numerical implementation, even if one adopts the same progenitor model, EOS, and treatment of gravity. In particular, numerical damping, which depends on the reconstruction method and the Riemann solver, may affect the oscillation modes in simulations. To see such a dependence on the numerical scheme, we also consider the results discussed in \cite{STT2021}, which was obtained with effective GR using the S20 progenitor model and the SFHo EOS, i.e., with the same setup for the physics as the other \textsc{3DnSNe} simulation in this study. Hereafter, the results shown in \cite{STT2021} will be denoted as STT21. In contrast to our new \textsc{3DnSNe} simulation, a different reconstruction scheme has been adopted in STT21, i.e., 2nd-order reconstruction with van Leer limiter combined with the  Harten-Lax-van Leer contact (HLLC) Riemann solver.

By comparing simulations with effective GR and GR in the xCFC approximation, different reconstruction schemes, and different progenitor models, we are able to investigate which form of mode relations is most suitable and robust for capturing the mode frequencies in supernova simulations.

It is useful to also compare the dynamical evolution of the various supernova models to ascertain that the simulations are similar enough, in general, to allow for a meaningful comparison of the gravitational wave emission. Moreover, only a limited number of code comparisons have been conducted so far for multi-dimensional supernova simulations \cite{just18,varma21}. Besides our main purpose of studying proto-neutron star oscillation modes, our model set also serves the sensitivity of supernova dynamics and proto-neutron star structure to the numerics and physics assumed in the models as a secondary purpose.

\begin{figure}[tbp]
\begin{center}
\includegraphics[scale=0.5]{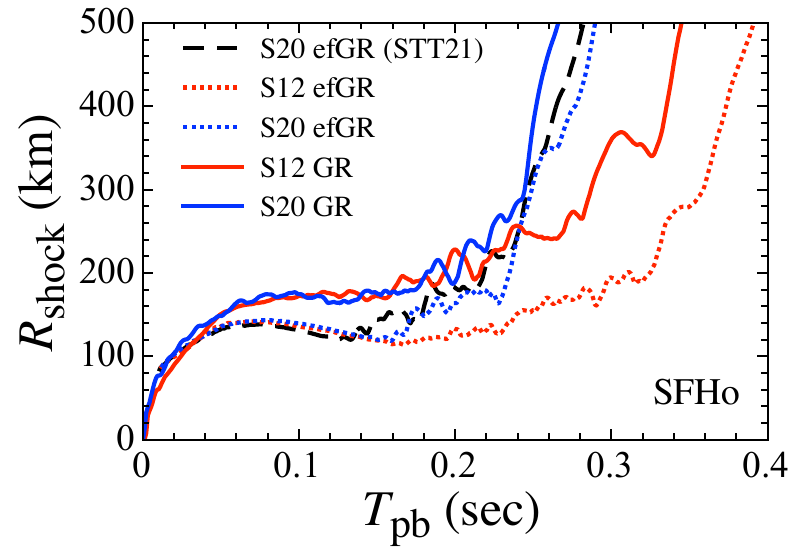} 
\end{center}
\caption{
Evolution of the shock radius for the supernova models considered in this study. The dotted and solid lines denote the results with effective GR and GR treatments, while the dashed line denotes the results shown in STT21 with effective GR but a different numerical scheme (see the text for details).
}
\label{fig:Rshock}
\end{figure}

In Fig.~\ref{fig:Rshock}, we show the time dependence of the shock radius for the supernova models considered in this study. First, one can observe that the behavior with effective GR using the S20 model is very similar to that obtained in STT21, i.e., the behavior of explosion may be less sensitive to the reconstruction scheme in performing the simulations. Moreover, one can observe that the S20 model explodes more quickly than the S12 model, both for the simulations with effective GR and the GR ones. The GR \textsc{CoCoNuT-FMT} models are characterized by somewhat larger shock radii than the  \textsc{3DnSNe} models prior to the explosion. However, the explosion develops at a similar time for S20, and the shock trajectories during the explosion phase are similar. For S12, shock revival in the \textsc{3DnSNe} model is delayed more visibly, but the shock radius still crosses $400\,\mathrm{km}$ at a similar time (within $50\,\mathrm{km}$ of each other).

\subsection{Proto-neutron star properties}
\label{sec:PNS}

For making a linear analysis, we have to prepare a spherically symmetric background proto-neutron star model. So, using the simulation data shown above, the properties such as pressure, density, sound velocity, and so on, are averaged in the angular direction. Unlike a cold neutron star, the surface of a proto-neutron star is not a sharp boundary, i.e., the matter exists even outside the proto-neutron star. In this study, as usual, we set the proto-neutron star surface where the density becomes $10^{11}$~g/cm$^3$.

\begin{figure}[tbp]
\begin{center}
\includegraphics[scale=0.5]{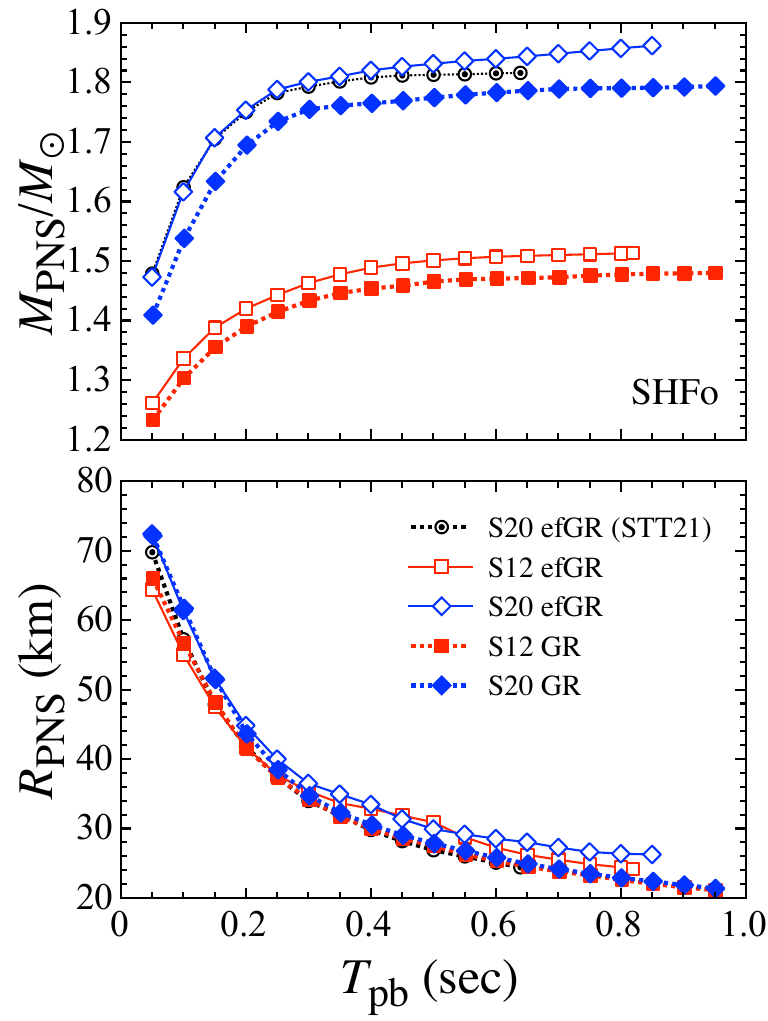} 
\end{center}
\caption{
The time evolution of the PNS mass (top) and radius (bottom), $T_{\rm pb}$ denotes the core bounce time. The surface density is chosen to be  $10^{11}$ g/cm$^3$. The open marks denote the results with effective GR, while the filled marks denote those with GR. For reference, the results with effective GR discussed in STT21 are also shown by double-circles. 
}
\label{fig:MRt}
\end{figure}

In general, the proto-neutron star mass increases with time due to the mass accretion, while the radius decreases due to the relativistic effect. In Fig.~\ref{fig:MRt}, we show the mass and radius of the proto-neutron stars considered in this study as a function of the core bounce time, $T_{\rm pb}$, where the open and filled marks denote the results with effective GR and GR, respectively. For reference, we also show the results shown in STT21 with double-circles.

We note that the GR models with \textsc{CoCoNuT-FMT} show slightly lower proto-neutron star masses, indicative of small structural differences in the proto-neutron star. Such small differences can arise, e.g., because the dynamics of the collapse phase is not perfectly identical, which is also reflected in different bounce times. Furthermore, one can see that the proto-neutron star radius with GR shrinks faster than that with effective GR. The radii are in close agreement in the first $0.2\, \mathrm{s}$ after the bounce, but then start to diverge more markedly. We also find that the proto-neutron star mass and radius slightly depend on the numerical scheme by comparing the result with effective GR using the S20 model to the result in STT21, even though the behavior of shock radius is very similar to each other as shown in Fig.~\ref{fig:Rshock}. Second-order results in a slightly smaller final proto-neutron star mass in the STT21 simulation (with second-order reconstruction) for model S20 compared to the new model with fifth-order reconstruction.

\begin{figure}[tbp]
\begin{center}
\includegraphics[scale=0.5]{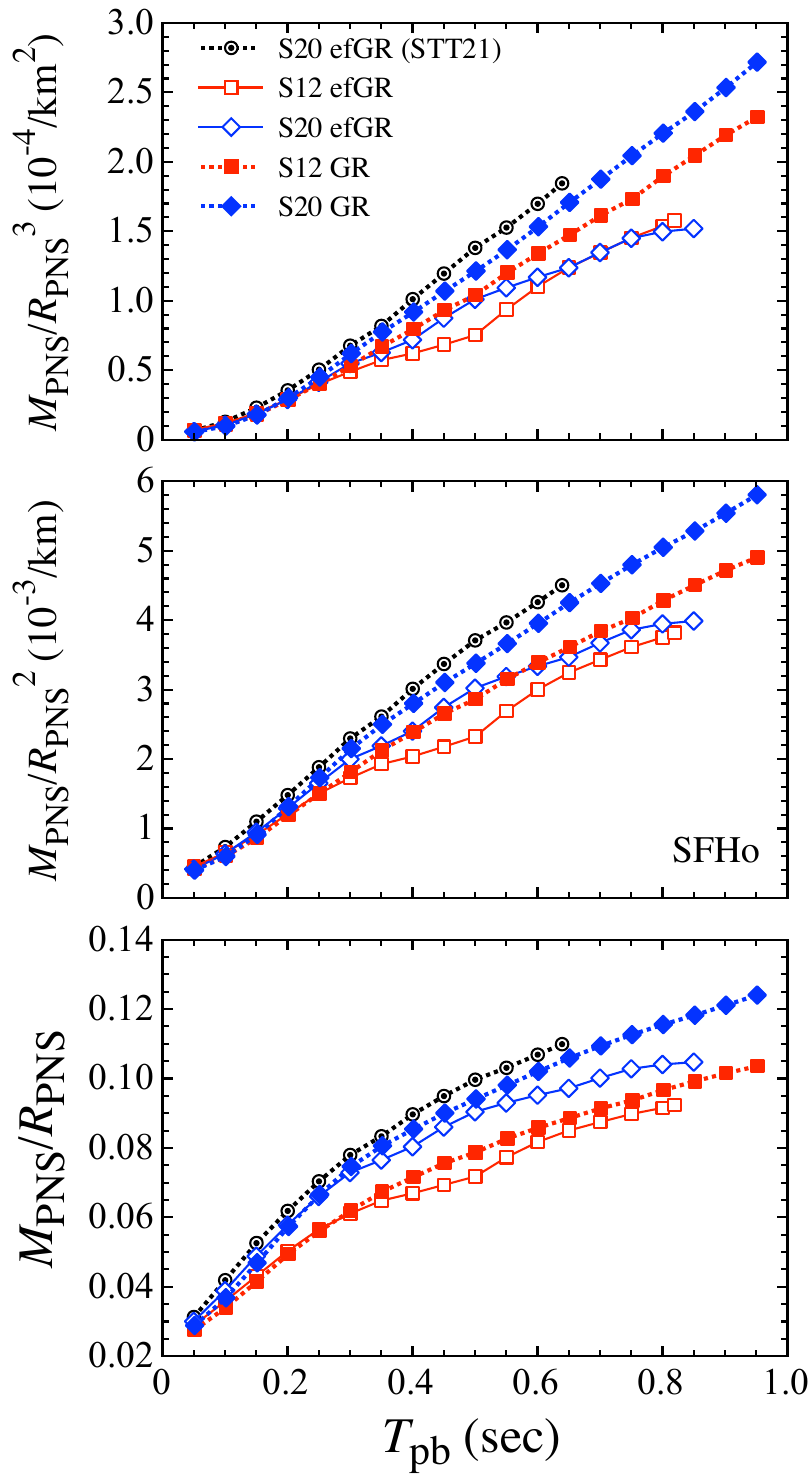} 
\end{center}
\caption{
Same as in Fig.~\ref{fig:MRt}, but for the stellar average density (top panel), the surface gravity (middle panel), and compactness (bottom panel).
}
\label{fig:MR3t2tt}
\end{figure}

In Fig.~\ref{fig:MR3t2tt}, we show the time evolution of the average density (top panel), (Newtonian) surface gravity (middle panel), and compactness (bottom panel) for the various proto-neutron star models, which are important properties for discussing the universal relations for the characteristic gravitational wave frequencies. From this figure, one can see a small deviation of the results in STT21 from those calculated in this study with effective GR using the S20 model, i.e., the time evolution of $M_{\rm PNS}/R_{\rm PNS}^3$, $M_{\rm PSN}/R_{\rm PNS}^2$, and $M_{\rm PNS}/R_{\rm PNS}$ are slightly sensitive to the numerical scheme in the simulation. The differences between GR and effective GR are more pronounced, especially after $200\, \mathrm{ms}$ post-bounce. This is mainly due to the divergence of the proto-neutron star radii and is particularly noteworthy in the average density (since the radius enters into this quantity as the third power).

Furthermore, as pointed out in \cite{ST2020a,STT2021}, we can confirm that the surface gravity is tightly correlated with the average proto-neutron star density, almost independently of the stellar progenitor model (and the numerical scheme used in performing the simulations) as shown in Fig.~\ref{fig:MR3_MR2}. The physical origin of this strong correlation yet has not been explained so far, but we explain the likely reason below. Since one can see the small, but systematic deviation in the results with GR by carefully observing Fig.~\ref{fig:MR3_MR2}, the correlation between the surface gravity and the average density may be associated with the treatment of gravity.

\begin{figure}[tbp]
\begin{center}
\includegraphics[scale=0.5]{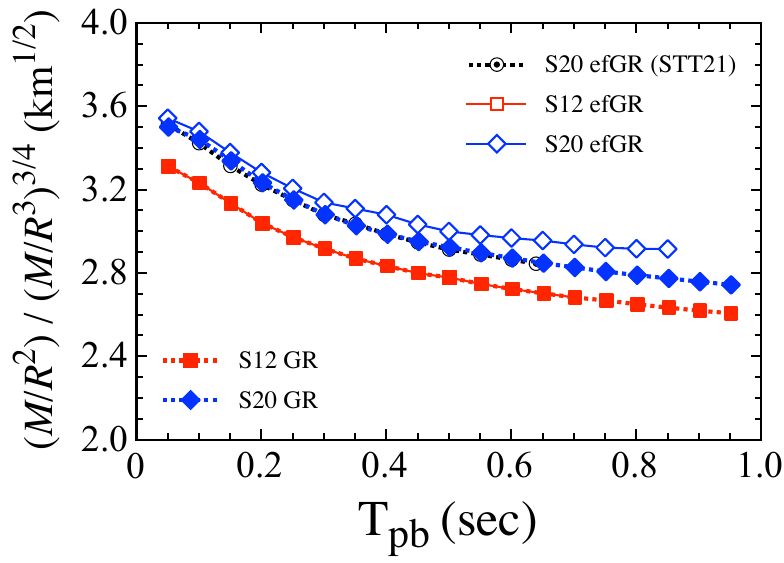} 
\end{center}
\caption{
Evolution of the value of $(M/R^2)/(M/R^3)^{3/4}$ for various proto-neutron star models.  
}
\label{fig:MR}
\end{figure}

\begin{figure}[tbp]
\begin{center}
\includegraphics[scale=0.5]{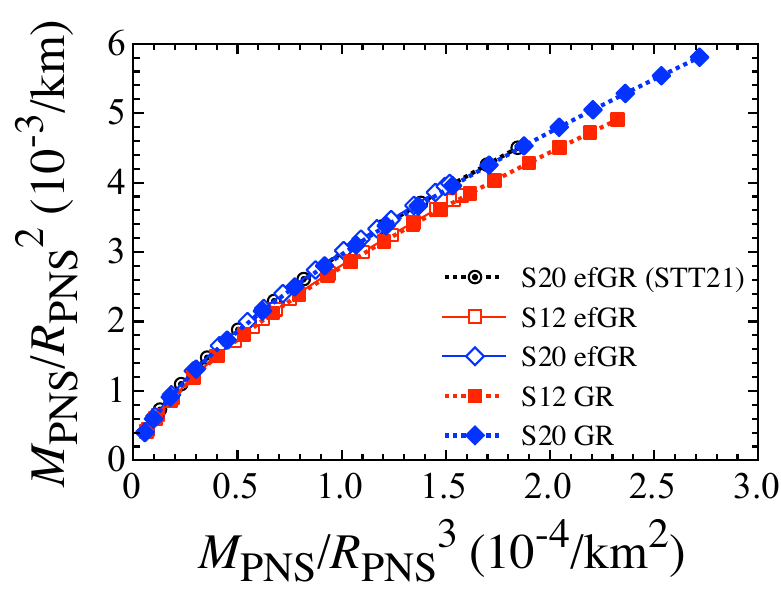} 
\end{center}
\caption{
Relation between the stellar average density and surface gravity. 
}
\label{fig:MR3_MR2}
\end{figure}

The correlation between the average density and the surface gravity can be explained by the proto-neutron star contraction law. During the few hundred milliseconds after the bounce, the contraction of the gain radius $r_\mathrm{g}$ approximately follows the relation $r_\mathrm{g} \propto \dot{M}^{1/3} M^{-1}$ in terms of the proto-neutron star mass $M$ and the accretion rate $\dot M$ \cite{mueller2016}. The ratio between the neutron star radius $R_\mathrm{PNS}$ and the gain radius is fairly constant, and hence a similar relation holds between $R_\mathrm{PNS}$ and $M$. Therefore the surface gravity and the average density scale as
\begin{align}
    \frac{M_\mathrm{PNS}}{R_\mathrm{PNS}^2} &\propto
    \frac{M_\mathrm{PNS}^3}{\dot{M}^{2/3}},\\
    \frac{M_\mathrm{PNS}}{R_\mathrm{PNS}^3} &\propto
    \frac{M_\mathrm{PNS}^4}{\dot{M}}.
\end{align}
This leads to
\begin{equation}
     \frac{M_\mathrm{PNS}}{R_\mathrm{PNS}^2} \propto
     \left(\frac{M_\mathrm{PNS}}{R_\mathrm{PNS}^3}\right)^{3/4}\dot{M}^{1/12},
\end{equation}
i.e., a power-law dependence between $M_\mathrm{PNS}/R_\mathrm{PNS}^2$, $M_\mathrm{PNS}/R_\mathrm{PNS}^3$ and the accretion rate $\dot{M}$. As the dependence on the accretion rate is weak (see Fig.~\ref{fig:MR}), this effectively amounts to a tight relation between surface gravity and average density. The dependence on the accretion rate in fact explains the slightly higher values of $M_\mathrm{PNS}/R_\mathrm{PNS}^2$ for the S20 simulations.

\section{Gravitational wave asteroseismology}
\label{sec:GW}

Using the proto-neutron star models discussed in the previous section as input, we perform a linear eigenmode analysis. We adopt the relativistic Cowling approximation, i.e., the metric perturbations are neglected when formulating the eigenvalue problem for the fluid oscillations. The perturbation equations are derived from the linearized energy-momentum conservation law. In formulating the oscillation equations one has to impose boundary conditions at the stellar center and surface. We impose the regularity condition at the center and the condition that the Lagrangian perturbation of pressure should be zero at the proto-neutron star surface. The concrete form of the perturbation equations and boundary conditions are the same as shown in \cite{SKTK2019}. Finally, adopting the normalized condition somewhere inside the star, the problem to solve becomes an eigenvalue problem with respect to the eigenvalue, $\omega$. The eigenfrequencies of the proto-neutron star at each time step are then given by $f=\omega/(2\pi)$. We note that the $f$-mode frequencies, which correspond to the gravitational wave signal in the simulations, determined with the full metric perturbations or even with parts of metric perturbations become lower than those determined with the Cowling approximation, if one adopts the same background proto-neutron star models \cite{ST2020c}.
Actually, the $g$-mode ($p$-mode) frequencies with Cowling approximation tend to be underestimated (overestimated), compared to the same modes determined with full metric perturbations.

\begin{figure}[tbp]
\begin{center}
\includegraphics[scale=0.5]{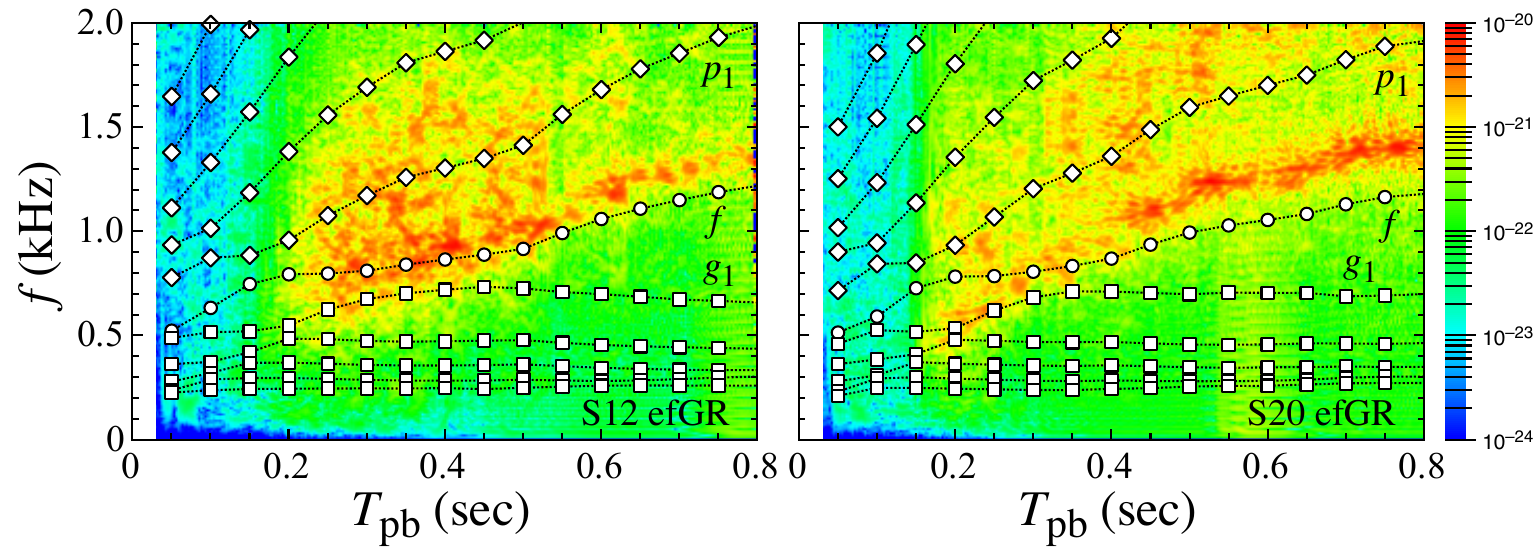} 
\end{center}
\caption{
Comparison of the gravitational wave signals obtained from the numerical simulations with effective GR (background contour) with the specific proto-neutron star oscillations (open marks), where the left and right panels correspond to the results with the S12 and S20 models. The circles, squares, and diamonds denote the $f$-, $g_i$-, and $p_i$-mode frequencies for $i=1$--$5$.
}
\label{fig:GW_efGR}
\end{figure}

We first consider the results with effective GR. In Fig.~\ref{fig:GW_efGR} we compare the gravitational wave spectrograms from the numerical simulations (background contour) and the specific oscillation frequencies of the proto-neutron star (marks), where we plot the $f$-, $g_i$-, and $p_i$-modes for $i=1-5$ with circles, squares, and diamonds, respectively, where the spectrograms is obtained by short-time Fourier transforms (see Ref. \cite{ST2020b} for detail). The left and right panels show the results for the S12 and S20 models. As discussed in our previous studies, the gravitational wave emission bands in the simulations seem to correspond to the $f$-mode ($g_1$-mode) oscillations of the proto-neutron star in the later (earlier) phase after (before) the avoided crossing between the $f$- and $g_1$-modes. However, as also mentioned in STT21, the frequencies of proto-neutron stars calculated with the Cowling approximation are systematically lower than the gravitational wave signals in the simulations done with effective GR (see Fig.~9 in STT21). This feature is consistent with the previous results, i.e., the presence of a systematic deviation is not sensitive to the numerical scheme. However, the deviation of $f$-mode frequencies from the gravitational wave signals in the simulation with the S20 model newly done in this study becomes significantly larger than the other models with the S12 model and the results in STT21. For example, the deviation at $T_{\rm pb}=0.6$ sec is $\sim 100$ Hz for STT21 and S12 model, while it becomes $\sim 200$ Hz for S20 model.

\begin{figure}[tbp]
\begin{center}
\includegraphics[scale=0.5]{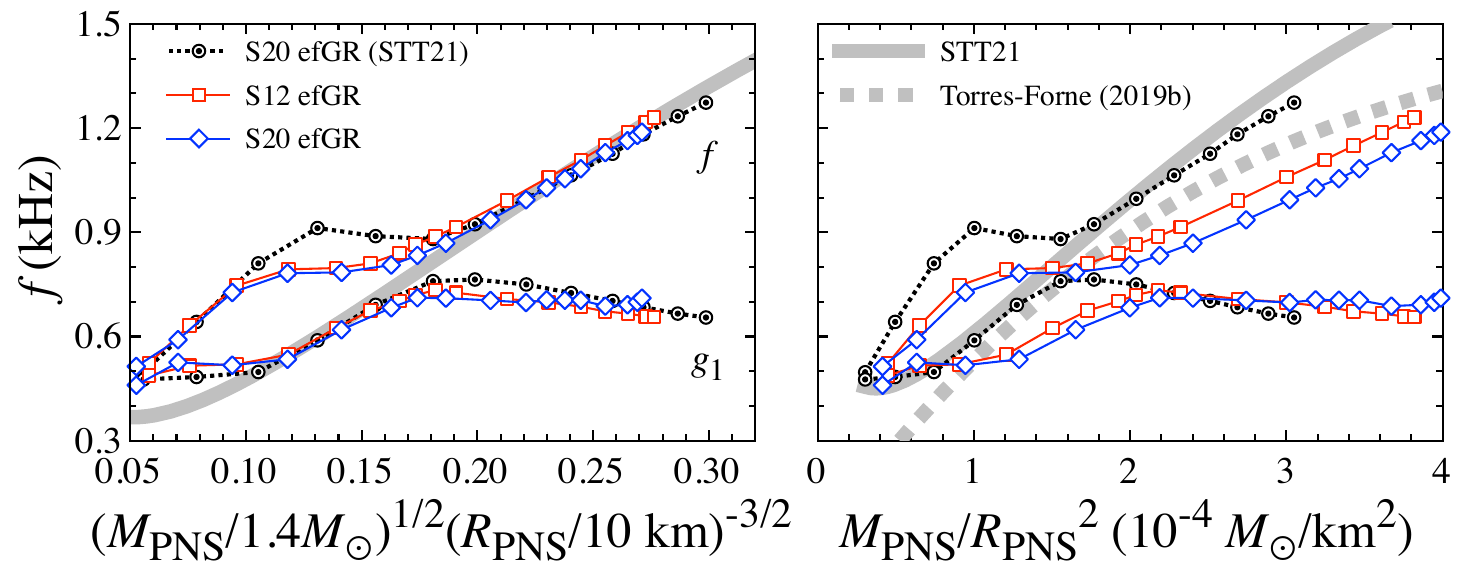} 
\end{center}
\caption{
The $f$- and $g_1$-mode frequencies on the evolving proto-neutron stars with effective GR are shown as a function of the normalized stellar average density (left panel) and surface gravity (right panel). The open squares and open diamonds correspond to the results with the S12 and S20 models, while the double circles denote the results shown in STT21 with the S20 model. The thick solid lines denote the empirical formulae derived in STT21, while the dotted line in the right panel is that derived in~\cite{TCPOF2019b}.
}
\label{fig:Universal-efGR}
\end{figure}

Next, we investigate the match of different proposed universal frequency relations from the literature
to the actual dominant gravitational wave frequency $f_\mathrm{GW}$. We consider the dependence on the square root of the average proto-neutron star density proposed by STT21, adopting 5 different supernova models obtained by 2D simulations in effective GR, 
\begin{equation}
    f_\mathrm{GW}\propto \left(\frac{M_\mathrm{PNS}}{R_\mathrm{PNS}^3}\right)^{1/2},
\end{equation}
and the dependence on surface gravity proposed by \cite{TCPOF2019b},
\begin{equation}
    f_\mathrm{GW}\propto \frac{M_\mathrm{PNS}}{R_\mathrm{PNS}^2}.
\end{equation}
We note that the empirical relation with the average proto-neutron star density has been derived with the eigenfrequencies determined with the Colwing approximation, while the relation with the surface gravity is with the eigenfrequencies determined with \textsc{GREAT} \cite{TCPOF2019a}, which includes parts of the metric perturbations. In Fig.~\ref{fig:Universal-efGR}, the frequencies of the $f$- and $g_1$-mode oscillations are shown as a function of the root square of the average density (left panel) and the surface gravity of the proto-neutron stars (right panel). The solid lines denote the empirical relations derived in STT21, while the dotted line denotes that derived in~\cite{TCPOF2019b}. From this figure, one can see that the actual trajectory of the dominant frequency of the gravitational wave signals from the branches of the $g_1$- to $f$-modes is best expressed with the empirical relation as a function of the root square of the average density. Interestingly, the $f$-mode frequency somewhat depends on the numerical scheme before the avoided crossing, but not after the crossing when $f_\mathrm{GW}$ follows the $f$-mode. For the $g_1$-mode the opposite is true. Intriguingly, the reconstruction method in the hydrodynamics solver seems to matter less when either of the two modes is actually the dominant emitter of gravitational waves.

By contrast, the frequencies of the $g_1$- to $f$ modes depend noticeably on the numerical scheme if we plot them as a function of surface gravity. In fact, in the right panel, one can observe that the results obtained in this study significantly deviate from our empirical relation as a function of the surface gravity (solid line), but they are more or less expressed with the empirical relation by ~\cite{TCPOF2019b} (dotted line). However, the scatter around the proposed scaling with surface gravity is clearly much bigger than for the density-frequency relation.
We thus conclude that the average density is a more suitable property for universally expressing the sequence from the $g_1$- to $f$-mode frequencies instead of the surface gravity.

\begin{figure}[tbp]
\begin{center}
\includegraphics[scale=0.5]{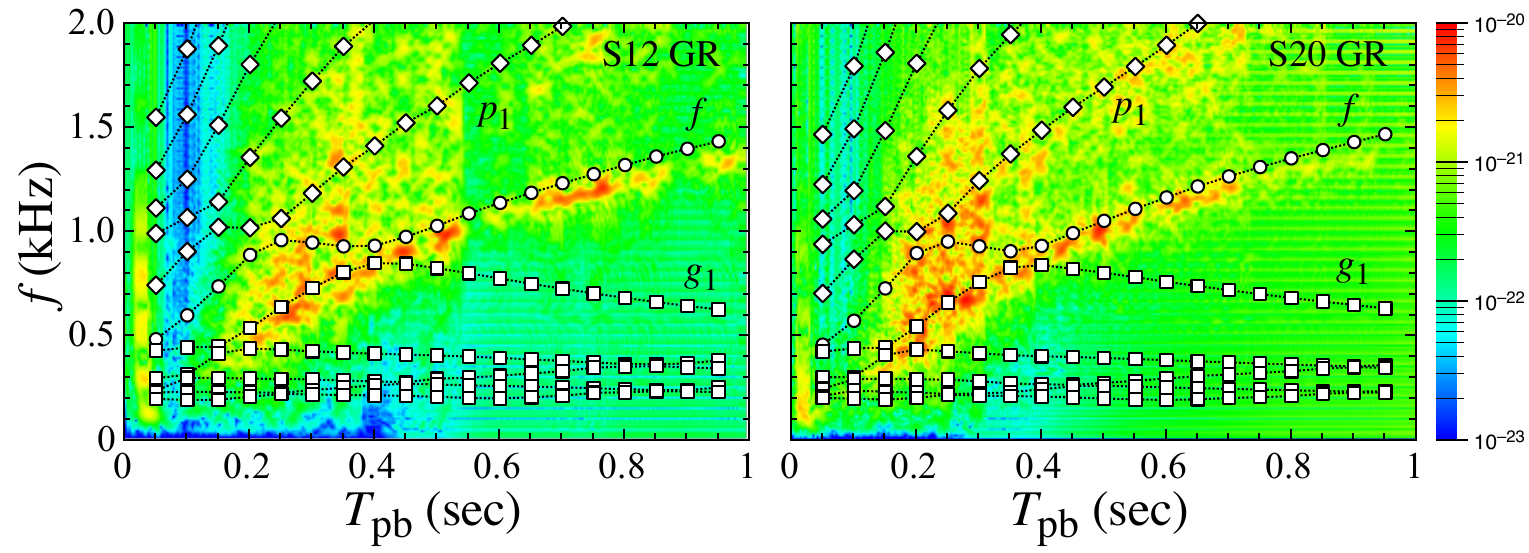} 
\end{center}
\caption{
Comparison of the gravitational wave signals obtained from the numerical simulations with GR (background contour) with the specific proto-neutron star oscillations (open marks), where the left and right panels correspond to the results with the S12 and S20 models. The circles, squares, and diamonds denote the $f$-, $g_i$-, and $p_i$-mode frequencies for $i=1-5$.
}
\label{fig:GW_GR}
\end{figure}

Next, we consider the effect of the treatment of gravity. Similar to Fig.~\ref{fig:GW_efGR}, we compare the specific frequencies of proto-neutron star oscillations (open marks) with the gravitational wave spectrograms from the numerical simulation with GR (background contour). Again one can see a good agreement between the linear mode analysis and the actual signals from the simulations. But, if this result is carefully compared to the results with effective GR shown in Fig.~\ref{fig:GW_efGR}, we find that the resultant frequencies of proto-neutron star oscillations slightly become larger than the frequencies of the gravitational wave signals in the simulations with GR, which is an opposite feature in the results with effective GR.

\begin{figure}[tbp]
\begin{center}
\includegraphics[scale=0.5]{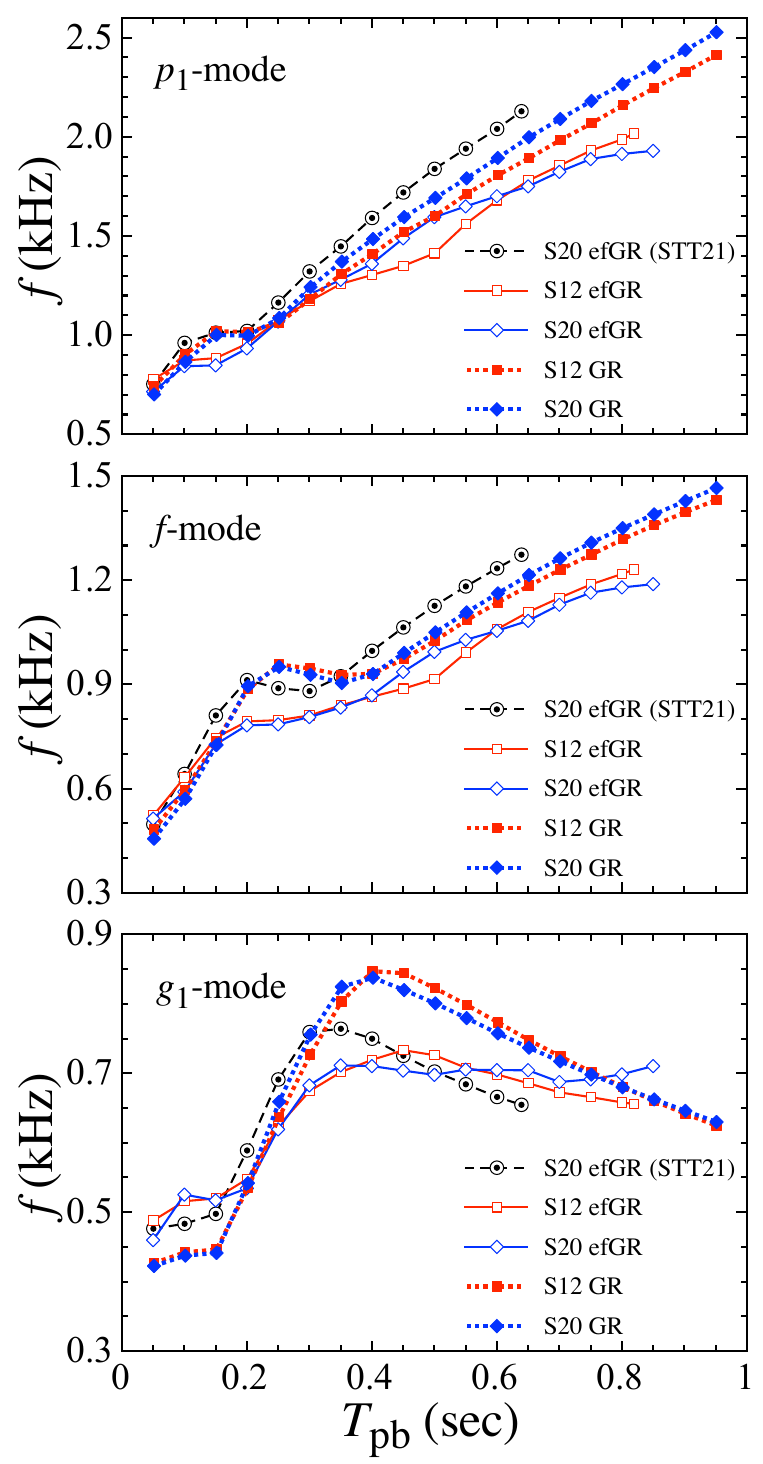} 
\end{center}
\caption{
Comparison of the time evolution of the proto-neutron star oscillation frequencies (the $p_1$-, $f$-, $g_1$-modes from top to bottom panels), using the data of numerical simulations with effective GR (solid lines) and with GR (dotted lines). The squares and diamonds denote the results with the S12 and S20 models. For reference, the results in STT21 are also shown with the double-circles. 
}
\label{fig:ft}
\end{figure}

It is interesting to consider to what extent the GR treatment and other details of the numerical implementation affect the trajectories of the mode frequencies. In Fig.~\ref{fig:ft}, we therefore compare the time evolution of the proto-neutron star oscillation frequencies obtained with the numerical simulations with effective GR (solid lines) and with GR (dotted lines), where the squares and diamonds denote the results with the S12 and S20 models. For reference, the results in STT21 are also shown as double-circles. From this figure, one can observe that the GR treatment and other details of the numerical implementation (such as the reconstruction scheme and the different neutrino transport in \textsc{3DnSNe} and \textsc{CoCoNuT}) introduce significant deviations in the evolution of the proto-neutron star frequencies. These variations tend to be larger than the progenitor dependence, especially for the $f$- and $g$-modes.

\begin{figure}[tbp]
\begin{center}
\includegraphics[scale=0.5]{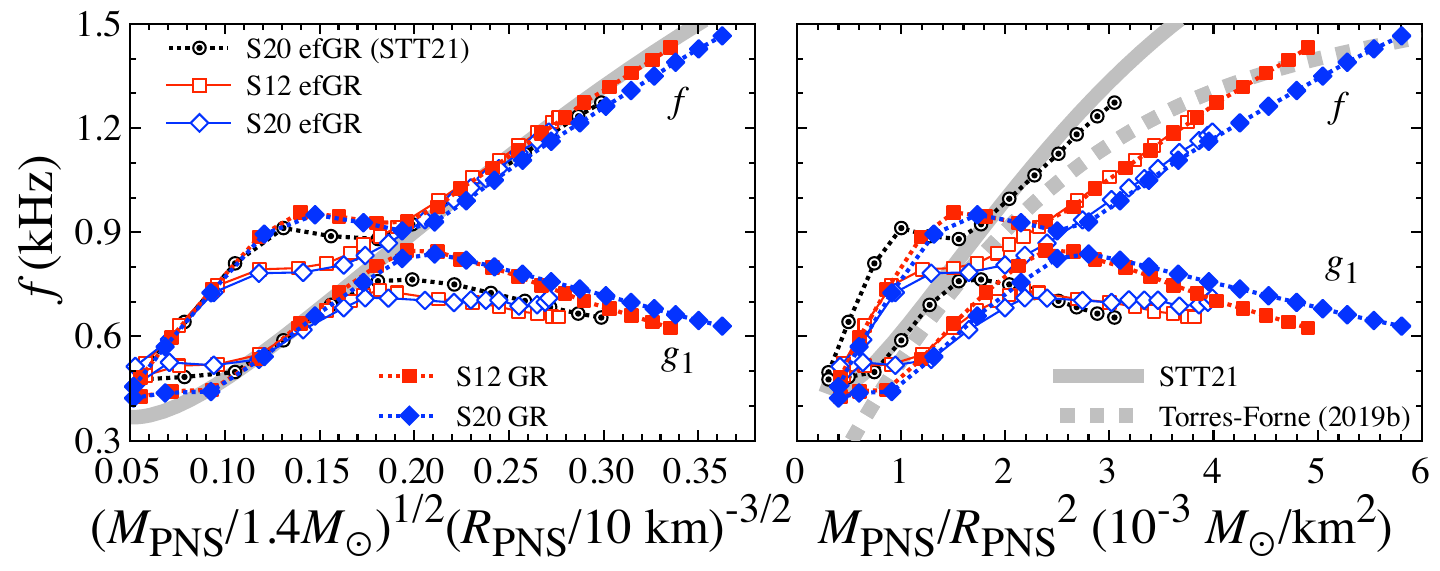} 
\end{center}
\caption{
Same as Fig.~\ref{fig:Universal-efGR}, but we also add the results with GR. The filled squares and diamonds correspond to the results with the S12 and S20 models with GR. 
}
\label{fig:Universal-GR}
\end{figure}

However, these differences in the frequency trajectories are largely a result of different proto-neutron star structures in the simulations. The uncertainties related to the numerical implementation prove much less significant when one tries to directly relate mode frequencies to proto-neutron star parameters.

To illustrate this, we show the $g_1$- and $f$-mode frequencies of the proto-neutron stars depend on the proto-neutron star properties both for the GR simulations and the effective GR simulations in Fig.~\ref{fig:Universal-GR}. Again we plot the mode frequencies both as a function of the square root of the normalized average density (left panel) and surface gravity (right panel). We observe that the sequence from the $g_1$- to $f$-mode frequencies corresponding to the gravitational wave signals in the simulations are well expressed with the proto-neutron star average density, even for the case with GR simulations, with very little scatter between the GR and the effective GR models. Again, the variations between the simulations are much bigger where the $g_1$- to $f$-mode happen not to be the dominant emission mode. On the trajectory of the dominant frequency, as a consequence, the GR treatment affects the avoided crossing between the $f$ and $g_1$-modes (which may be observable as an emission gap in the spectrum~\cite{MRBV2018}) with GR seems to occur with more massive and dense proto-neutron star models than that with effective GR. We also find that the dependence of the $g_1$- and $f$-modes on the surface gravity with GR simulation are more or less similar to those with effective GR simulations newly done in this study.

We conclude that, independently of the treatment of gravity in the simulations or other differences in the numerical implementation, the average density always appears a more suitable predictor for the $f$-mode frequencies excited in proto-neutron stars determined with the Cowling approximation than the surface gravity. On the other hand, how well the frequencies of proto-neutron stars match the gravitational wave signal in the numerical calculations depends on the treatment of gravity in the simulations, even though the dominant gravitational wave signals in the simulations seem to be identified with the $f$-mode ($g_1$-mode) frequencies of the proto-neutron stars in the later (earlier) phase after the core bounce. Nevertheless, this statement may still depend on the supernova models, because we obtained the conclusion from a few models. The comparison of the proto-neutron star frequencies with the fitting formula with the surface gravity is beginning to be made, e.g., the top panel of Fig.~7 in \cite{mori23}. Similar to our discussion, some models deviate from the fitting formula (see Fig.~8 in \cite{wolfe23}). The scatter from the formula is evaluated in Ref.~\cite{Bizouard21}. On the other hand, the fitting formula with the average density has not been extensively compared to oscillation frequencies of proto-neutron stars obtained from other supernova models yet, except for Ref.~\cite{mori23}. Although the horizontal axis of the bottom panel of Fig.~7 in Ref.~\cite{mori23} seems to contain some error, their proto-neutron star frequencies deviate from our fitting formula with the average density, 
where the simulation has been done in \textsc{GR1D} using the progenitor model with $9.6M_\odot$ and zero-initial metallicity. This deviation may come from the light progenitor mass, the simulation in one dimension, and/or the long-simulation time, i.e., $20$ seconds postbounce.
So, it may be important to verify our conclusion in this study, using various models \cite[e.g.,][]{Warren20,Bizouard21,wolfe23,Vartanyan23,Bruel23}. 
In addition, we showed that the universal relation discussed here is independent of the EOSs at least for the EOSs we adopted here, i.e., DD2, SFHo, Togashi, and LS220. But, to confirm the EOS independence of our relation, one may additionally have to check the proto-neutron star models with other EOSs.

\begin{figure}[tbp]
\begin{center}
\includegraphics[scale=0.5]{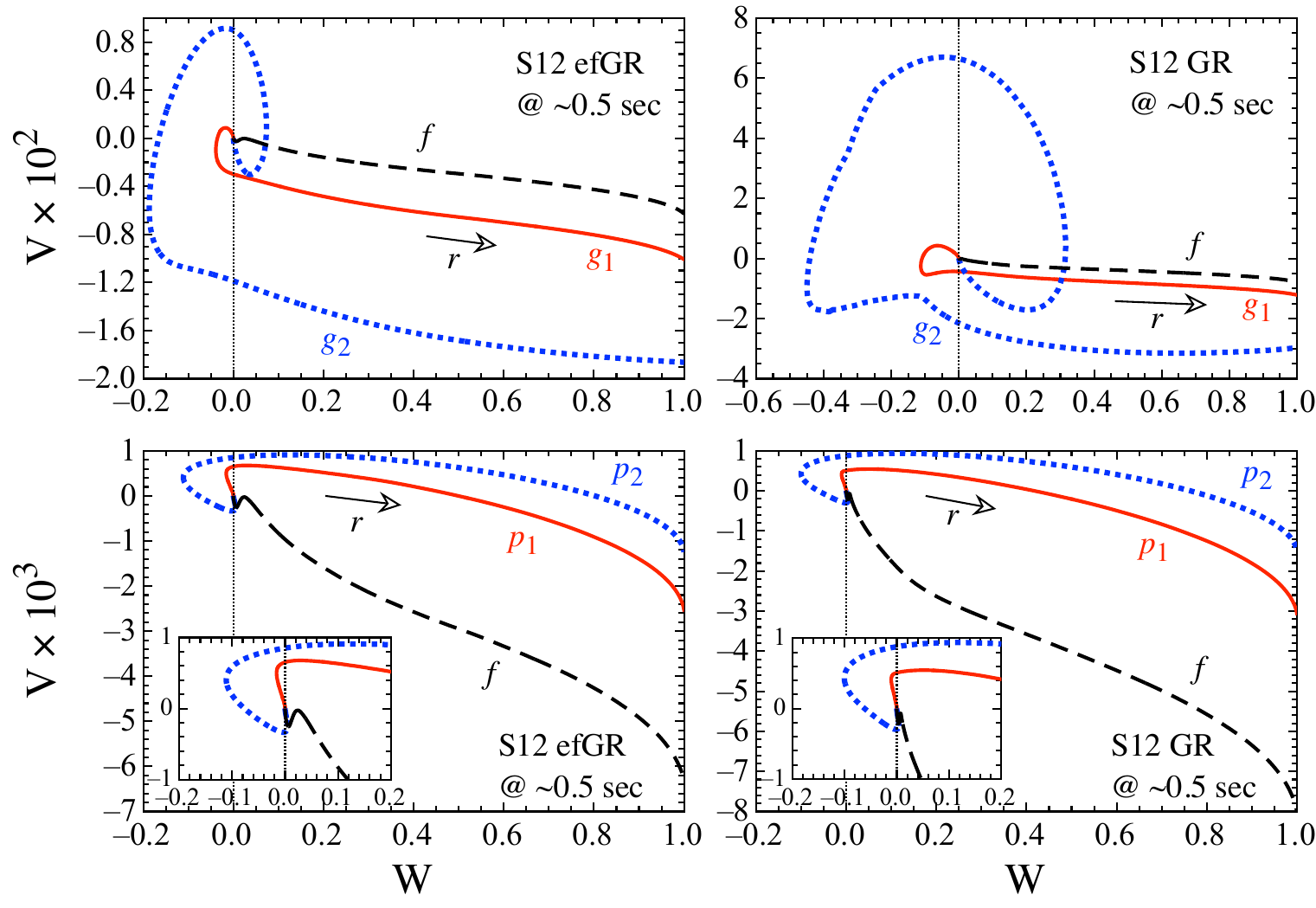} 
\end{center}
\caption{
Phase diagram for the $g_1$- and $g_2$-modes in the top panels, while $p_1$- and $p_2$-modes in the bottom panels for the protoneutron stars at $T_{pb}\sim 0.5$ sec. The left and right panels correspond to the results with effective GR and GR, respectively. 
}
\label{fig:class}
\end{figure}

Finally, we have comments on the mode classification. We simply identify the oscillation modes by counting the nodal number of the eigenfunctions in this study, as in our previous studies, e.g., Fig. 3 in Ref. \cite{ST2020b}, while one may also identify the modes by checking the behavior of the eigenfunctions in phase diagram \cite{TCPF2018,rodriguez23}. That is, if one plots the displacement in the radial direction, $W$, and in the angular direction, $V$, as the radial coordinate, $r$, increases, the trajectory rotates counterclockwise for a gravity wave (or $g$-mode-like oscillations) and clockwise for a sound wave (or $p$-mode-like oscillations). We note that, since the definition of our angular displacement is an opposite sign (See Ref. \cite{SKTK2019}) compared to Refs. \cite{TCPF2018,rodriguez23}, the rotational direction is also reversed. In practice, as shown in Fig. \ref{fig:class}, the $g$-modes ($p$-modes) rotate counterclockwise (clockwise) in the phase diagram, where the top (bottom) panels correspond to the $g$-modes ($p$-modes), while the left (right) panels correspond to the case with effective GR (GR) at $T_{pb}\sim 0.5$ sec. In Fig.~\ref{fig:class}, we also show the $f$-mode for reference, and $W=0$ with the thin-dotted (vertical) line.

\begin{figure}[tbp]
\begin{center}
\includegraphics[scale=0.5]{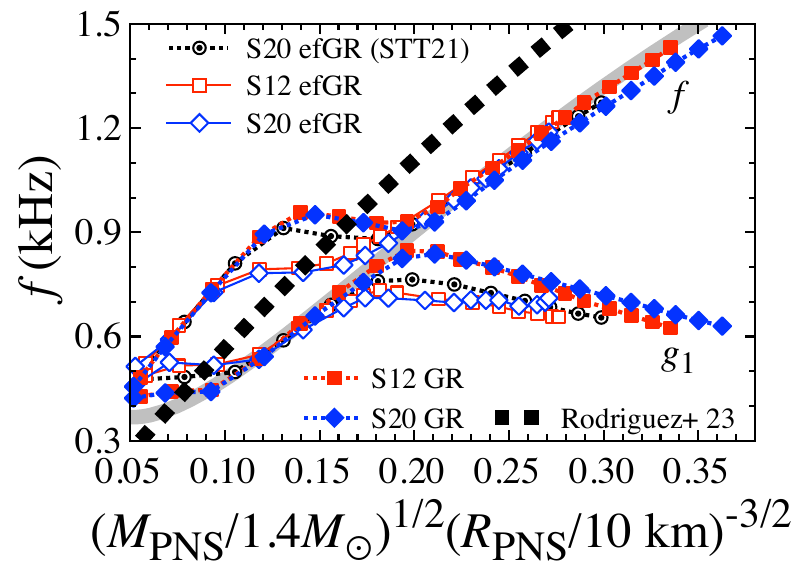} 
\end{center}
\caption{
Comparison of the universal relation (thick-dotted line) derived in Ref. \cite{rodriguez23} to the results obtained in this study shown in the left panel of Fig. \ref{fig:Universal-GR}. 
}
\label{fig:compare}
\end{figure}

In addition to the standard mode classification, another classification based on modal property can be considered. Adopting this classification, it has been reported that the avoided crossing appearing in the standard classification disappears \cite{TCPF2018}, and the universal relations of the gravitational frequencies with proto-neutron star properties, e.g., the stellar average density, seem to extremely become tight \cite{rodriguez23}. In fact, the universal relations of the gravitational wave frequencies with the classification based on the modal property have been derived as a function of stellar average density \cite{rodriguez23}, even though the unit of the fitting coefficients listed in Table 2 in Ref. \cite{rodriguez23} might be kHz instead of Hz. In Fig. \ref{fig:compare}, we compare their universal relations to our results obtained in this study. The reason why the universal relation in \cite{rodriguez23} deviates from our results even though their relation is also as a function of the stellar average density, maybe the difference in how to construct the proto-neutron star models. That is, in Ref. \cite{rodriguez23} they reconstructed the proto-neutron star models by solving the Tolman-Oppenheimer-Volkoff equations as Ref. \cite{ST2016}, using the three-dimensional simulation data, while our models are prepared directly from the two-dimensional simulation data by averaging in the angular direction. In addition to the difference in the proto-neutron star models, the metric perturbation has been partially taken into account in Ref. \cite{rodriguez23}, while we simply calculate the frequencies with the Cowling approximation. 
Anyway, it is important to stress that our study shows, quite remarkably, that the same universal relation in terms of average density holds in two different codes independent of the treatment of gravity (modified potential vs. conformal flatness condition), which is non-trivial.

\section{Conclusion}
\label{sec:Conclusion}

The supernova gravitational waves are one of the most promising candidates next to the gravitational waves from the compact binary mergers. To extract the physical properties using the observed gravitational wave signals, a kind of supernova-model-independent universal relation(s) between the signal and proto-neutron star properties is quite useful, if exists, because the spectrogram of gravitational waves strongly depends on the supernova models. Up to now, the universal relations expressing the $g_1$- and $f$-mode frequencies of the proto-neutron stars, which correspond to the gravitational wave signals in numerical simulations, as a function of average density or surface gravity of the proto-neutron stars have mainly been discussed. In this study, to see which proto-neutron star property is more suitable to universally express the gravitational wave signals, we examine the specific oscillation frequencies of the proto-neutron stars, using the numerical simulation with effectively relativistic (effective GR) and relativistic (GR) treatment of gravity. In addition to these examinations, to see the dependence on the interpolations in the simulations, we also compare the results with the previous results shown in \cite{STT2021}. Then, we find that the $f$-mode frequencies of the proto-neutron stars determined with the Cowling approximation are estimated smaller (larger) than the gravitational wave signals in numerical simulations with effective GR (GR) treatment. Furthermore, we find that the sequence from the $g_1$- to $f$-mode frequencies is expressed as a function of the average density almost independently of the progenitor mass, the treatment of gravity, and the interpolations in the simulations. On the other hand, the relation between the frequencies of the corresponding sequence and surface gravity depends on the treatment of gravity and especially the interpolations in the simulations. Therefore, the average density must be more suitable to universally express the supernova gravitational waves rather than the surface gravity of the proto-neutron stars, at least in our models for the with $\sim 1\,\mathrm{s}$ of the post-bounce phase.
In this study, we discuss the universal relation for the frequencies of proto-neutron star oscillations determined with the Cowling approximation, but one can expect that a similar universal relation may exist even without the Cowling approximation. We will examine such a possibility in the future.

\acknowledgments

This work is partly supported in part by Japan Society for the Promotion of Science (JSPS) KAKENHI Grant Number (
JP19KK0354,  
JP21H01088,  
JP22H01223,  
JP23K03400,  
and JP23K20848). 
This research is also supported by FY2023 RIKEN Incentive Research Project,
and by Pioneering Program of RIKEN for Evolution of Matter in the Universe (r-EMU),
and MEXT as “Program for Promoting 
researches on the Supercomputer Fugaku” (Structure and Evolution of the Universe Unraveled by Fusion of Simulation and AI; Grant Number JPMXP1020230406) and JICFuS.
Part of numerical computations were carried out on PC cluster at Center for Computational Astrophysics, National Astronomical Observatory of Japan.



\end{document}